\documentclass[preprint,aps]{revtex4}

\begin{document}
\title{An analytic expression for the electronic correlation term of the kinetic functional.}
\author{Luigi Delle Site}
 \email{dellsite@mpip-mainz.mpg.de}
\affiliation{Max-Planck-Institute for Polymer Research\\
Ackermannweg 10, D 55021 Mainz Germany.}

\begin{abstract}
We propose an analytic formula for the non-local Fisher information
functional, or electronic kinetic correlation term, appearing in the expression of the kinetic
density functional. Such an explicit formula is constructed on the basis of well founded physical arguments and a
rigorous mathematical prescription.  
\end{abstract}
\maketitle
\section{introduction}
In a previous work we have treated the problem of a rigorous determination of the
upper bound for the kinetic density functional of a system of electrons\cite{jpamio}. There, starting from the most
general expression of a many-body wavefunction in polar form, a final formula
was obtained where the kinetic functional consisted of two terms. The first,
linked to the phase factor of the wavefunction, we have shown, can be
bounded from above by a term proportional to the Thomas-Fermi one, the
second corresponds to the Weizsacker term. A third term, corresponding to
what is known as the non-local Fisher information functional, was neglected
since, 
being quantitatively much smaller than the other terms \cite{lud0}, it was not relevant for the discussion carried on there.
However, although quantitatively not relevant, such a term  
plays a fundamental role regarding important structural properties, such as the atomic shell structure,
and it is strictly related to the electronic correlations implicit in the kinetic
functional \cite{ludrev} (for this reason we have renamed it as ''kinetic 
correlation term'').
The difficulty in treating it is due to the fact that an explicit
rigorous expression does not exists and to find a suitable one represents a
challenging task
\cite{kohout}. To our knowledge, only few attempts have been made in this
direction and they were restricted to highly simplified situations \cite{lud1}.
Clearly, for a detailed electronic description of condensed matter systems, 
this terms may play a key role. For this reason we focus the attention on such
a problem and propose a general procedure, physically well founded and
mathematically rigorous, to obtain an analytic expression. 
The practical importance of this problems stems from the fact that, as it is
discussed in \cite{jpamio}, there is a renewed interest towards
 energy functionals which are nowadays used in flexible quantum (based)
 computational methods,
such as the linear-scaling real-space kinetic energy functional method, where the kinetic energy is calculated as a functional of the electron density. Such a method is also known as Orbital Free Density Functional (OFDFT) (see e.g.\cite{watson}).
These modern methods allow in principle for the treatment of much larger systems and
longer time scales than those based on standard Density Functional
Theory in the Kohn-Sham formulation \cite{yangparr}. The crucial point of
OFDFT, is that it is in principle exact, provided that the kinetic functional
is rigorous. This means that the designing of rigorous kinetic density
functionals would allow to explore the full power
of the Hohenberg-Kohn theorem for extended systems. In the light of the
discussion above, the search for a rigorous kinetic functional is not simply 
interesting for its own theoretical implications and understanding of
fundamental aspect of quantum mechanics, but also for its large importance in
practical calculations in condensed matter. The paper is organized as follows:
the many-body wavefunction approach to obtain a
$3N$-dimensional functional (with $N$, the number of electrons), which
was presented in our previous
work, is reviewed and extended to the case where the electrons are explicitly
interacting. Next the reduction to a $3$-dimensional case for each term
appearing in the $3N$-dimensional expression is treated. While the term
related to the wavefunction phase factor and the Coulomb interaction 
are known, the part related to the kinetic correlation is new and represent
the core of this work.
\section{The $N$-particle wavefunction approach for the Hamiltonian problem} 
Let us consider the many body Hamiltonian operator for a system of $N$ interacting electrons:
\begin{equation}
H=-\frac{1}{2}\sum_{i=1}^{N}\nabla^{2}_{i}+\frac{1}{2}\sum_{j>i}\frac{1}{|{\bf r}_{i}-{\bf r}_{j}|}
\end{equation}
where for simplicity the Planck constant $\hbar$, the electron mass $m$ and the electron
charge $e$ are taken to be 1.
Let us consider the most general form of the wavefunction solution of the stationary Schr\"{o}dinger equation for an arbitrary state of energy $\epsilon$:
\begin{equation}
H\psi=\epsilon \psi
\label{eq1}
\end{equation} 
Such a general form can be written as:
\begin{equation} 
\psi({\bf
  r}_{1},......,{\bf r}_{N})=e^{iS({\bf r}_{1}......,{\bf r}_{N})}
\Theta({\bf r}_{1},.....,{\bf r}_{N})
\end{equation}
with $S$ and $\Theta$ real functions
in an $N$-dimensional space of volume $\Omega=\Pi_{i=1}^{N}\omega_{i}=\omega^{N}$, with $\omega$
being the single particle volume. We also require $S$ to be symmetric and $\Theta$ to be antisymmetric with respect to any particle pair
permutation. $\psi$ satisfies the normalization condition:
$\int_{\Omega}\psi^{*}\psi d{\bf r}_{1}...,d{\bf r}_{N}=1$.
The  spin variables are not explicitly considered, however, their effects  will be properly taken into account and specifically mentioned when it will be required.
Our aim is to use Eq.\ref{eq1} and the form and properties of $\psi$ to derive
an energy functional $E[\rho]=\int F[\rho({\bf r})]d{\bf r}$ where $
F[\rho({\bf r})]$ is the energy density written as a functional of the one
particle electron density, $\rho({\bf
  r})$,  defined as:
\begin{eqnarray} 
\rho({\bf r})=\rho({\bf
r}_{1})= N\int_{\Omega_{N-1}} \psi^{*}({\bf r}_{1},....,{\bf r}_{N})\psi({\bf r}_{1},....,{\bf r}_{N})d{\bf
r}_{2}......d{\bf r}_{N}
\end{eqnarray}
where $\Omega_{N-1}$
is the $N-1$-dimensional volume of $N-1$ particles, and $\int\rho({\bf r})d{\bf r}=N$; since the particles are
indistinguishable, the index $1$ can be exchanged with any other of the
remaining $N-1$ particle index, for the rest of the work this formalism will
be kept. 
$E[\rho]$ is the equivalent of $\epsilon$ in
Eq.\ref{eq1} written in terms of $\rho({\bf r})$; thus we should obtain: 
\begin{equation}
E[\psi]=\int_{\Omega}(\psi^{*}H\psi) d{\bf r}_{1}....d{\bf r}_{N}=\epsilon=E[\rho]=\int F[\rho({\bf r})]d{\bf r}.
\label{eq2}
\end{equation}
To this aim we consider the $N$-particle energy density ${\hat F}_{\psi}=\psi^{*}H\psi$; the explicit expression is \cite{jpamio}:
\begin{eqnarray}
{\hat F}_{\psi}& = & \frac{1}{2}\sum_{i=1,N}\Theta^{2}({\bf r}_{1},..,{\bf r}_{N})|\nabla_{i}S({\bf r}_{1},..,{\bf
  r}_{N})|^{2}+\frac{1}{2}\sum_{i=1,N}|\nabla_{i} \Theta({\bf r}_{1},..,{\bf r}_{N})|^{2}+{} \nonumber\\ && {}+\frac{1}{2}\sum_{i=1}^{N}\sum_{j\neq
  i}^{1,N}\frac{\Theta^{2}({\bf r}_{1},..,{\bf r}_{N})}{|{\bf r}_{i}-{\bf r}_{j}|}-\frac{1}{4}\sum_{i=1,N}\nabla_{i}^{2}\Theta^{2}({\bf r}_{1},..,{\bf r}_{N}).
\label{eq3}
\end{eqnarray}
Eq.\ref{eq3} is defined in a $3N$-dimensional space; in order to derive
 $F[\rho({\bf r})]$, we shall reduce the problem to a $3$-dimensional space. Note that the last term on the r.h.s. vanishes after integration, thus it will not be relevant in this context.
The first step to reduce the problem to a $3$-dimensional space, is to redefine $\Theta^{2}({\bf r}_{1},.....{\bf
 r}_{N})$ in terms of a one particle probability density and an $(N-1)$-particle
 conditional probability density:
\begin{equation}
\Theta^{2}({\bf r}_{1},....,{\bf r}_{N})=\frac{\rho({\bf r}_{1})}{N}f({\bf
  r}_{2},.......,{\bf r}_{N}/{\bf r}_{1})
\label{eq4}
\end{equation}
where $=\frac{\rho({\bf r}_{1})}{N}$ is the one particle probability density
 and $f({\bf
  r}_{2},.......,{\bf r}_{N}/{\bf r}_{1})$ is the
$N-1$ electron conditional (w.r.t. ${\bf r}_{1}$) probability density, i.e. the
probability density of finding an $N-1$ electron configuration, $C({\bf
  r}_{2},.......,{\bf r}_{N})$, for a given fixed
value of ${\bf r}_{1}$. The function $f$ satisfies the following property:
\begin{equation}
\int_{\Omega_{N-1}}f({\bf
  r}_{2},........,{\bf r}_{N}/{\bf r}_{1})d{\bf
  r}_{2}.......d{\bf r}_{N}=1 \forall {\bf
  r}_{1}
\label{condeq1}
\end{equation}
 as before, due to the electron indistinguishability,
the index $1$ was chosen arbitrarily. Another equivalent factorization of
$\Theta^{2}$ , which it will be used for simplifying the electron-electron
Coulomb term, consists of writing $\Theta^{2}$ as a product of 
a two particle probability density:
\begin{equation}
 \frac{\rho({\bf r}_{1},{\bf r}_{2})}{N}=\int_{\Omega_{N-2}} \psi^{*}({\bf
   r}_{1},{\bf r}_{2},{\bf r}_{3}....,{\bf r}_{N})\psi({\bf r}_{1},{\bf
   r}_{2},{\bf r}_{3}....,{\bf r}_{N})d{\bf
r}_{3}......d{\bf r}_{N}
\end{equation}
equivalent to the diagonal element of the second-order spinless density matrix \cite{yangparr},
and an $N-2$ particle conditional probability density:
 \begin{equation}
f({\bf
  r}_{3},.......,{\bf r}_{N}/{\bf r}_{1},{\bf r}_{2}).
\end{equation}
Equivalently to the factorization shown before, in this case
we have (see e.g. \cite{ayers}):
\begin{equation}
\int \rho({\bf r}_{1},{\bf r}_{2})d{\bf r}_{1}d{\bf r}_{2}=N
\end{equation} 
and:
\begin{eqnarray}
\int_{\Omega_{N-2}}f({\bf
  r}_{3},.....,{\bf r}_{N}/{\bf r}_{1},{\bf r}_{2})
  d{\bf r}_{3}....d{\bf r}_{N}=1 \forall {\bf r}_{1},{\bf r}_{2}.
\end{eqnarray}
As a consequence of the definitions given above, the following identity is satisfied:
\begin{eqnarray} 
\Theta^{2}({\bf r}_{1},...{\bf r}_{N})& =& \frac{\rho({\bf r}_{1})}{N}f({\bf
  r}_{2},..,{\bf r}_{N}/{\bf r}_{1})={} \nonumber\\ && {}=
\frac{\rho({\bf r}_{1},{\bf r}_{2})}{N}f({\bf
  r}_{3},..,{\bf r}_{N}/{\bf r}_{1},{\bf r}_{2}). 
\label{eq5}
\end{eqnarray}
Once again, since the particles are indistinguishable, the indices $1$ and
  $2$ for the one and two particle density were chosen arbitrarily and any other choice would
  be equivalent. 
Inserting Eq.\ref{eq4} and the r.h.s. of Eq.\ref{eq5} into Eq.\ref{eq3}, using the properties of $f$ shown above, and remembering that the particles are
  indistinguishable, we obtain (for the kinetic terms see also \cite{sears,jpamio}):
\begin{eqnarray}
\int_{\Omega}(\psi^{*}H\psi)d{\bf r}_{1}...d{\bf
  r}_{N}= \frac{1}{2}\int_{\Omega}\sum_{i=1}^{N}\Theta^{2}({\bf r}_{1},...,{\bf r}_{N})|\nabla_{i}
S({\bf r}_{1},....,{\bf
  r}_{N})|^{2}d{\bf r}_{1}...d{\bf r}_{N}+\frac{1}{8}\int_{\omega}\frac{|\nabla\rho({\bf
    r}_{1})|^{2}}{\rho({\bf r}_{1})}d{\bf
  r}_{1}+{}\nonumber\\ {}+ \frac{1}{8}\int_{\omega}\rho({\bf r}_{1})\left[\int_{\Omega_{N-1}}\frac{|\nabla_{{\bf r}_{1}}f({\bf r}_{2},....,{\bf r}_{N}/{\bf r}_{1})|^{2}}{f({\bf r}_{2},.....,{\bf r}_{N}/{\bf r}_{1})}d{\bf r}_{2}....d{\bf
  r}_{N}\right]d{\bf r}_{1}+&&\nonumber\\ +\frac{(N-1)}{2}\int_{\omega}\int_{\omega}\frac{\rho({\bf r}_{1},{\bf
      r}_{2})}{|{\bf r}_{1}-{\bf r}_{2}|}d{\bf r}_{1}d{\bf
    r}_{2}.
\label{eq6}
\end{eqnarray}
The last term on the r.h.s. of Eq.\ref{eq6} is obtained in the following way: first we use the property of
particle indistinguishability and rewrite the Coulomb term in Eq.\ref{eq3} as :
\begin{equation}
\frac{1}{2}\sum_{i=1}^{N}\sum_{j\neq i}^{1,N}\int_{\Omega}\frac{|\Theta({\bf
    r}_{1},....,{\bf r}_{N})|^{2}}{|{\bf r}_{i}-{\bf r}_{j}|} d{\bf
    r}_{1}...d{\bf r}_{N}=\frac{N(N-1)}{2}\int_{\Omega}\frac{|\Theta({\bf
    r}_{1},....,{\bf r}_{N})|^{2}}{|{\bf r}_{1}-{\bf r}_{2}|} d{\bf
    r}_{1}...d{\bf r}_{N}
\label{eq7}
\end{equation}
where the indices $1$ and $2$ are, as usual, chosen arbitrarily, then we use the decomposition of 
$\Theta^{2}$ as a product of the two particle density $\rho({\bf r}_{1},{\bf
  r}_{2})$, and the two particle conditional probability density, $f({\bf
  r}_{3},....,{\bf r}_{N}/{\bf r}_{1},{\bf r}_{2})$, and obtain:
\begin{eqnarray}
\frac{N(N-1)}{2}\int_{\Omega}\frac{|\Theta({\bf
    r}_{1},....,{\bf r}_{N})|^{2}}{|{\bf r}_{1}-{\bf r}_{2}|} d{\bf
    r}_{1}...d{\bf r}_{N}=\frac{(N-1)}{2}\int_{\omega}\int_{\omega}\frac{\rho({\bf
    r}_{1},{\bf r}_{2})}{|{\bf r}_{1}-{\bf r}_{2}|}d{\bf r}_{1}d{\bf r}_{2}\nonumber\\ \int_{\Omega_{N-2}}f({\bf
    r}_{3},....,{\bf r}_{N}/{\bf r}_{1},{\bf r}_{2})d{\bf r}_{3}.....d{\bf r}_{N}
\label{eq8}
\end{eqnarray}
Eq.\ref{eq8} can be simplified to a two particle expression, since 
\begin{eqnarray}
\int_{\Omega_{N-2}}f({\bf
    r}_{3},......,{\bf r}_{N}/{\bf r}_{1},{\bf r}_{2}) d{\bf r}_{3}.....d{\bf
    r}_{N}=1 \forall {\bf r}_{1},{\bf r}_{2}.
\end{eqnarray}
so that we have:
\begin{equation}
\frac{N(N-1)}{2}\int_{\Omega}\frac{|\Theta({\bf
    r}_{1},....,{\bf r}_{N})|^{2}}{|{\bf r}_{1}-{\bf r}_{2}|} d{\bf
    r}_{1}...d{\bf r}_{N}=\frac{(N-1)}{2}\int_{\omega}\int_{\omega}\frac{\rho({\bf r}_{1},{\bf r}_{2})}{|{\bf r}_{1}-{\bf r}_{2}|}d{\bf r}_{1}d{\bf r}_{2}
\end{equation}
\section{From the $N$-particle to the one particle representation}
As one can see from Eq.\ref{eq6} there are two kinetic terms which are written in  
an $N$-particle representation, namely, the phase factor term: 
\begin{equation}
\int_{\Omega}\frac{1}{2}\sum_{i=1}^{N}\Theta^{2}({\bf r}_{1},...,{\bf r}_{N})|\nabla_{i}
S({\bf r}_{1},....,{\bf r}_{N})|^{2}
\end{equation}
and the correlation term also known as the Fisher non-local information functional
\cite{fisher,sears,nagy,romera}:
\begin{eqnarray}
\frac{1}{8}\int_{\omega}\rho({\bf r}_{1})d{\bf
  r}_{1}\int_{\Omega_{N-1}}\frac{|\nabla_{{\bf r}_{1}}f({\bf r}_{2},.....,{\bf r}_{N}/{\bf r}_{1})|^{2}}{f({\bf r}_{2},.....,{\bf r}_{N})} d{\bf r}_{2}.......d{\bf r}_{N}
\label{fisheq}
\end{eqnarray}
and the aim of this work is to reduce them to a one-particle
representation; the first term was already considered in a previous work
\cite{jpamio} and the results will be reported here for completeness; the
second term deserves a particular attention, since to our knowledge only very
few attempts, restricted to specific and simple cases \cite{lud1}, were made
to give an explicit formula; we will propose a rather general procedure to
obtain an analytic expression.
The third term, the
electron-electron Coulomb interaction, which, as we have seen before, is a two
particle term, and, generalizing the index $1$ and $2$, can be
written as:
\begin{equation}
E_{ee}=\frac{(N-1)}{2}\int_{\omega}\int_{\omega'}\frac{\rho({\bf r},{\bf r'})}{|{\bf r}-{\bf r'}|}d{\bf r}d{\bf r'}
\end{equation}
it has been extensively treated in
literature and the related results will play an important role in the
procedure of reducing the $N$-particle expression of Eq.\ref{fisheq} to a single particle functional expression.   
\subsection{Reduction of $\int _{\Omega}\sum_{i=1,n}\Theta^{2}|\nabla_{i} S|^{2}d{\bf
    r}_{1}...{\bf r}_{N}$ to a one particle representation}
In our previous work \cite{jpamio} we have shown that in the limit of a slowly varying density we can reasonably conjecture the following inequality:
\begin{equation}
|\nabla_{i}S|\leq |{\bf P}_{F_{i}}|
\label{eqmom}
\end{equation}
which means that the momentum of the $i$-th particle is less (or equal) to the Fermi momentum $P_{F_{i}}$ assigned to the $i$-th particle (i.e. the maximum value of the momentum an electron, under such conditions, can have).
We have shown that Eq.\ref{eqmom} leads to the following inequality for the kinetic term related to the phase factor $S$:
\begin{equation}
\int_{\Omega}\sum_{i=1}^{N}\Theta^{2}({\bf r}_{1},...,{\bf r}_{N})|\nabla_{i}S({\bf r}_{1},...,{\bf r}_{N})|^{2}d{\bf r}_{1}...d{\bf r}_{N}\leq C\int_{\omega}\rho^{\frac{5}{3}}({\bf r})d{\bf r}.
\label{ineq}
\end{equation}
Where $C=\frac{(3\pi^{2})^{2/3}}{2}$.
We have also discussed the reasonability of our results beyond the slowly varying density regime and how our upper bound compares to others available in the literature. The effects of the spin are implicitly incorporated into the definition of ${\bf P}_{F_{i}}$, where the spin degeneracy enters as a scaling factor. On the other hand, any possible spin (or spatial) correlation effect implicit in $S$ can be neglected, because the energy assigned to each electron in the upper bound term is so large, or equivalently the overestimation of the energy is large enough, that any possible correlation energy can be reasonably neglected in Eq.\ref{ineq}. In this case the term ''reduction'' means that the $3N$-dimensional term is reduced to a $3$-dimensional term, expressed through the quantity $\rho({\bf r})$, which bounds the true expression from above. This is the first step in reducing $E[\psi]=\int_{\Omega}\psi^{*}H\psi d{\bf r}_{1}...d{\bf r}_{N}$ to a reasonable density functional. 
\subsection{The electron-electron interaction}
The electron-electron interaction energy
$E_{ee}=\frac{(N-1)}{2}\int_{\omega}\int_{\omega'}\frac{\rho({\bf
    r},{\bf r'})}{|{\bf r}-{\bf r'}|}d{\bf r}d{\bf r'}$, can be separated in two parts, the (classical ) Hartree term:
\begin{equation}
E_{H}=\int_{\omega}\int_{\omega'}\frac{\rho({\bf r})\rho({\bf r'})}{|{\bf r}-{\bf r'}|} d{\bf r}d{\bf r'}
\label{hatree}
\end{equation}
and the exchange and correlation term:
\begin{equation}
E_{xc}=\int_{\omega}\rho({\bf r})\epsilon_{xc}({\bf r})d{\bf r}
\label{exch}
\end{equation}
where $\epsilon_{xc}({\bf r})$ is the exchange and correlation energy density
(see e.g. \cite{yangparr}).
In order to describe the nature of the exchange and correlation term in a simple and compact way, we refer to a series of seminal papers by Wigner and his coworker Seitz \cite{wig,wigseit}.
In their work on the interaction of electrons in metals, they treat the case of uniformly distributed interacting electrons. They argue that the interaction energy of electrons with parallel spin is determined by the classical Coulomb interaction due to the space charge plus the exchange term. This latter, also known as Fermi correlation \cite{gosh}, is the consequence of the Pauli exclusion principle, and expresses the fact that two electrons with the same spin stay away from each other; such an effect is also called a Fermi hole. In contrast, the interaction of electrons with antiparallel spin contains the classical Coulomb term, plus the correlation term. The latter arises from the fact that electrons with antiparallel spin tend to condense in the same place but the electrostatic repulsion keeps them apart.  
We should also add that the exchange and correlation effects contained in
the electron-electron term, are, by now, well understood. Although their satisfactory description is by far not easy achievable, there is a rich
literature dealing with reasonable descriptions (numerical, analytical, 
semi-analytical) of such a quantity (see e.g. articles in \cite{lnp} and
references therein). It is not the purpose of this work to
propose anything new regarding this part; on the contrary, it is the
availability of various analytical formulas for the exchange and
correlation term which makes it possible to formally derive results on the kinetic
correlation term.
\subsection{The non-local Fisher Information Functional or Kinetic Correlation term}  
After having treated the kinetic term related to the wavefunction phase factor
and having given a general description of the direct electron-electron interaction,
we now come to the core of this work; to give a general prescription for
finding an analytical functional form for the non local Fisher information
functional or kinetic correlation term:
\begin{equation}
\frac{1}{8}\int_{\omega}\rho({\bf r}) I({\bf r})d{\bf r}=\frac{1}{8}\int_{\omega}\rho({\bf
  r})\left[\int_{\Omega_{N-1}}\frac{|\nabla_{\bf r}f({\bf r}_{2},....{\bf
  r}_{N}/{\bf r})|^{2}}{f({\bf r}_{2},....{\bf r}_{N}/{\bf r})}d{\bf
  r}_{2}...d{\bf r}_{N}\right]d{\bf r}.
\label{fishag}
\end{equation}
The main point is to give physically as well as mathematically well founded
conditions which, when satisfied, lead to a valid expression for $I({\bf r})$.
This term is qualitatively very important since it is responsible of the
atomic shell structure \cite{lud1,ludrev}, thus to have a valid description of it automatically
means to have an accurate description of the electronic
structure. Interestingly, Lude\~{n}a and Karasiev\cite{ludrev} intuitively link this term to the exchange and 
correlation energy, although a prescription for how to find this link is not
given. We start from the same intuitive idea and develop it further. The
essence of the idea developed here 
is that a gas of electrons, as long as these are not
related by an explicit interaction, does not display correlation effects,
except those of the basic Fermi distribution, or occupation of levels, for non
interacting electrons. However, when a direct electron-electron interaction is
switched on, then the electrons start to feel the presence of all the others
in a non trivial way. This means that their spatial distribution is heavily
governed by the fact that they try to optimize such correlations
(interactions)  in a way that the optimal spatial arrangement fulfills the
minimum energy requirements. For this reason the key quantity determining the
many particle electron distribution is the electron-electron interaction:
\begin{equation}
E_{ee}({\bf r}, {\bf r'})=\frac{\rho({\bf r},{\bf r'})}{|{\bf r}-{\bf
    r'}|}.
\label{ee}
\end{equation} 
The expression of Eq.\ref{ee} is a two-particle function, averaged over the
rest of $N-2$ particles, where the effects of these latter on the electrons at ${\bf r}$
and ${\bf r'}$ is implicitly described. As we have seen before,
although we do not have an exact expression for $E_{ee}({\bf r}, {\bf r'})$,
nevertheless several valid approximations are given in an analytic form.
Conversely, very little can be found in the literature for $I({\bf r})$ due
to the difficulty in writing an explicit expression for $f({\bf
  r}_{2},...,{\bf r}_{N}/{\bf r})$ \cite{kohout}. For this reason, 
since we cannot write directly $f({\bf
  r}_{2},...,{\bf r}_{N}/{\bf r})$ in an explicit way, we can think of 
transforming $f({\bf
  r}_{2},...,{\bf r}_{N}/{\bf r})$ into some function of $E_{ee}({\bf r}, {\bf
  r'})$; this approach, although indirect, can lead to a physically well-founded analytic expression. As said before,
$f({\bf r}_{2},...,{\bf r}_{N}/{\bf r})$ is the probability density of finding
$N-1$ particles in a certain configuration once one particle ${\bf r}$ is
fixed; it is reasonable to think that it must be linked to
the electron-electron interaction of Eq.\ref{ee}. For the reasons given above
we propose the following conjecture:
\begin{equation}
f({\bf r}_{2},...,{\bf r}_{N}/{\bf r})=\Pi_{i=2}^{N}h_{i}(E_{ee}({\bf r},{\bf
  r}_{i}))
\label{conj1}
\end{equation}      
where $h_{i}(E_{ee}({\bf r},{\bf r}_{i}))$ must satisfies the following conditions:
\begin{eqnarray}
(a)&\int_{\omega}h_{i}(E_{ee}({\bf r},{\bf r}_{i}))d{\bf r}_{i}=1;\forall {\bf r}
&\nonumber\\
\label{cond}
(b)& h_{i}(E_{ee}({\bf r},{\bf r}))=0 &\\
(c)& \Pi_{i=2}^{N}h_{i}(E_{ee}({\bf r},{\bf
  r}_{i}))\leq 1 ;\forall {\bf r},{\bf r}_{i}\nonumber&. 
\end{eqnarray} 
Condition $(a)$ comes directly from Eq.\ref{condeq1}, condition $(b)$ is the
direct consequence of the antisymmetry of $\psi$, while the last condition
expresses the fact that $f({\bf r}_{2},...,{\bf r}_{N}/{\bf r})$ is a
probability density, thus cannot be larger than one. The function
$\Pi_{i=2}^{N}h_{i}(E_{ee}({\bf r},{\bf r}_{i}))$ in this context should be
interpreted as the probability density which expresses how the ${\bf r}_{i}$ 
particles distribute themselves w.r.t. the ${\bf r}$-particle (fixed), while
experiencing each other (except ${\bf r}$) in an average way 
implicitly contained in $E_{ee}({\bf r}, {\bf r'})$ through $\rho({\bf r})$.
The conditions in Eq.\ref{cond} are general, and can be used to design
possible functional forms for $I({\bf r})$. In the next section we propose
some functional forms and discuss the consequences.
\subsubsection{The exponential form} 
In this section we will treat the case when the function
$h_{i}=h_{i}(E_{ee}({\bf r},{\bf r}_{i}))$ has the form of an exponential. By this
ansatz and its related mathematical prescription we will gain some interesting insights into, possibly, physical effects 
related to the kinetic correlation term we are treating. The functional form
we propose for $f$ has the following expression:
\begin{equation}
\Pi_{i=2}^{N}h_{i}(E_{ee}({\bf r},{\bf
  r}_{i}))=\Pi_{i=2}^{N}e^{(N-1){\overline E}({\bf r})}e^{-E_{ee}({\bf r},{\bf r}_{i})}
\label{exp1}
\end{equation}
where
\begin{equation}
e^{-{\overline E}({\bf r})}=\int_{\omega}e^{-E_{ee}({\bf r},{\bf
    r}_{i})}d{\bf  r}_{i}.
\label{norm}
\end{equation}
This function naturally satisfies the requirement $(a)$ and $(c)$ of Eq.\ref{cond}; in fact 
\begin{equation}
e^{{\overline E}({\bf r})}\times \int_{\omega}e^{-E_{ee}({\bf r},{\bf
      r}_{i})}d{\bf r}_{i}=1 ;\forall {\bf r}, i.
\end{equation}
and as a consequence, because of the particles' indistinguishability, we have:
\begin{equation}
 \int_{\Omega_{N-1}}\Pi_{i=2}^{N}e^{(N-1){\overline E}({\bf
 r})}e^{-E_{ee}({\bf r},{\bf r}_{i})}d{\bf r}_{2}...d{\bf r}_{N}=1
 ;\forall {\bf r}.
\end{equation}
Condition $(c)$ is satisfied, since $e^{{\overline E}({\bf r})}$ is the automatic normalization factor, and thus:
\begin{equation}
\frac{e^{-E_{ee}({\bf r},{\bf r}_{i})}}{e^{-{\overline E}({\bf r})}}\le 1;
  \forall {\bf r},{\bf r}_{i}
\label{min1}
\end{equation}
and again, this can be straightforwardly generalized  to the complete
  $\Pi_{i=2}^{N}h_{i}(E_{ee}({\bf r},{\bf
  r}_{i}))$ function.
We will come back to condition $(b)$ of Eq.\ref{cond} later on, for the moment
  let us justify why the exponential form should be physically sound.
The starting point is that the electron-electron interaction is a positive
  definite function, $E_{ee}({\bf r},{\bf r'})\ge 0;\forall {\bf r},{\bf r'}$.
The physical principle we use is that the electrons try to distribute themselves in a way that their spatial
  arrangement is consistent with the optimal
  electron-electron total interaction energy. In this context the conditional
  probability density, $f({\bf r}_{2},...{\bf r}_{N}/{\bf
  r})=\Pi_{i=2}^{N}\frac{e^{-E_{ee}({\bf r},{\bf r}_{i})}}{e^{-(N-1){\overline E}({\bf
  r})}}$ expresses the fact that for a configuration of high interaction the
  probability is low because electrons tend to avoid each other; on the other
  hand, for a lower interaction energy the probability is higher, because the
  electrons can distribute in a more optimal way. So far, on the basis of the
  physical intuition and mathematical requirements, our ansatz seems rather
  reasonable; however in this form, $f({\bf r}_{2},...{\bf r}_{N}/{\bf
  r})$ does not satisfy a major symmetry requirement, namely condition $(b)$
  of Eq.\ref{cond}. In fact $\Theta({\bf r}_{1},..{\bf r}_{N})$ being
  antisymmetric under any particle pair permutation, it would require that
$\Theta^{2}({\bf r},{\bf r}_{2},...{\bf r},...{\bf r}_{N})=0$.
We should remember that $\Theta^{2}({\bf r},{\bf r}_{2},...{\bf r},...{\bf
  r}_{N})=\rho({\bf r})f({\bf r}_{2},...{\bf r}_{N}/{\bf
  r})$, which implies that $f({\bf r}_{2},.,{\bf r},..,{\bf r}_{N}/{\bf
  r})=0$; this is impossible in the form we have expressed $f({\bf
  r}_{2},...{\bf r}_{N}/{\bf r})$. To understand this point we have to look at
  the explicit expression of $E_{ee}$:
\begin{equation}
E_{ee}({\bf r},{\bf r'})=\frac{\rho({\bf r},{\bf r'})}{|{\bf r}-{\bf
    r'}|}
\label{eeexp}
\end{equation}
thus:
\begin{equation}
\lim_{{\bf r}\to{\bf r'}}E_{ee}({\bf r},{\bf r'})=0
\label{eeexp1}
\end{equation}
i.e. the probability of finding two electrons in the same position is zero,
and thus the related interaction is zero too. This can be also seen, for example,
in Ref.\cite{gosh}, where $E_{ee}({\bf r},{\bf r'})$ is explicitly written in
term of classical Coulomb interaction plus the exchange and correlation term;
for ${\bf r}\to {\bf r'}$ the exchange and correlation part becomes equal to
minus the classical Coulomb term, and thus they cancel each other. For our
expression of $f$, instead, we must require $E_{ee}({\bf r},{\bf r'})\to
  \infty$ for ${\bf r'}\to{\bf r}$ so that $f({\bf r}_{2},.,{\bf r},..,{\bf
    r}_{N}/{\bf r})=0$, however this is not possible because $E_{ee}({\bf
      r},{\bf r})=0$ thus we cannot use $E_{ee}({\bf r},{\bf r'})$ as a
      central quantity. There is also another necessary condition which must
      be satisfied by $f({\bf r}_{2},..,{\bf
    r}_{N}/{\bf r})$, and it involves directly the kinetic correlation
  term; in fact, this latter must be finite whenever any one of the ${\bf r}_i$
electrons approaches the ${\bf r}$-electron; i.e.
\begin{equation}
\frac{|\nabla_{\bf r}f({\bf r}_{2},..,...{\bf r},...,{\bf
    r}_{N}/{\bf r})|^{2}}{f({\bf r}_{2},..,{\bf r},...,{\bf
    r}_{N}/{\bf r})}=finite
\label{finit}
\end{equation}
where $finite$ is in general either zero or a finite positive number or even a non diverging function of the other variables than ${\bf r}$. Eq.\ref{finit} implies that
the numerator should go to zero with the same order of, or faster than, the
denominator. This mathematical requirement together with condition
$(b)$ of Eq.\ref{cond} suggest that instead of the full electron-electron
interaction, $E_{ee}({\bf r},{\bf r'})$, we should use only the classical
Coulomb term, also known as the Hartree term, $E_{H}({\bf r},{\bf
  r'})=\frac{\rho({\bf r})\rho({\bf r'})}{|{\bf r}-{\bf r'}|}$. It can be
easily verified that it satisfies the conditions $(a)$, $(b)$, $(c)$ of
Eq.\ref{cond}, and leads to a finite value (zero) for the kinetic correlation
term in the limit of ${\bf r}_{i}\to {\bf r}$.
In fact in this case one has:
\begin{equation}
\frac{|\nabla_{{\bf r}}f({\bf r}_{2},..,{\bf
    r}_{N}/{\bf r})|^{2}}{f({\bf r}_{2},..,{\bf r}_{N}/{\bf r})}
    =\frac{\left|\nabla_{{\bf
    r}}\left[\Pi_{i=2}^{N}e^{(N-1){\overline E}({\bf r})}e^{-E_{H}({\bf r},{\bf
    r}_{i})}\right]\right|^{2}}{\left[\Pi_{i=2}^{N}e^{(N-1){\overline E}({\bf
    r})}e^{-E_{H}({\bf r},{\bf r}_{i})}\right]}.
\label{eqlim}
\end{equation}
The r.h.s. of Eq.\ref{eqlim} becomes:
\begin{equation}
e^{(N-1){\overline E}({\bf r})}\Pi_{i=2}^{N}e^{-E({\bf r},{\bf r}_{i})}\cdot \left|(N-1)\nabla_{{\bf r}}{\overline E}({\bf r})-\sum_{i=2}^{N}\nabla_{{\bf r}}E_{H}({\bf r},{\bf r}_{i})\right|^{2}
\label{mlim}
\end{equation}
and thus we have:
\begin{equation}
\lim_{{\bf r}_{i}\to{\bf r}}e^{(N-1){\overline E}({\bf r})}\Pi_{i=2}^{N}e^{-E({\bf r},{\bf r}_{i})}\cdot \left|(N-1)\nabla_{{\bf r}}{\overline E}({\bf r})-\sum_{i=2}^{N}\nabla_{{\bf r}}E_{H}({\bf r},{\bf r}_{i})\right|^{2}=0; \forall {\bf r}_{i}.
\label{finlim}
\end{equation}
The limit in Eq.\ref{finlim} is zero because while $\left|(N-1)\nabla_{{\bf r}}{\overline E}({\bf r})-\sum_{i=2}^{N}\nabla_{{\bf r}}E_{H}({\bf r},{\bf r}_{i})\right|^{2}$ diverges for ${\bf r}_{i}\to{\bf r}$, in the worst case, as $|{\bf r}-{\bf r}_{i}|^{-n}$ with $n$ a finite positive integer, the exponential $e^{-E_{H}({\bf r},{\bf r}_{i})}$ tends to zero in a much faster way than the divergent term.
The fact  that $I({\bf r})$, written using the exponential form of $f$, does not diverge for ${\bf r}_{i}\to {\bf r}$, together with the other mathematical properties discussed before, show that indeed the exponential ansatz is mathematically and physically sound, thus we can proceed to the core of this work, that is to write an explicit formula for $I({\bf r})$. 
From Eq.\ref{eqlim} and Eq.\ref{mlim} we obtain:
\begin{eqnarray}
I({\bf r})=(N-1)^{2}\left|\nabla_{\bf r}{\overline E}({\bf r})\right|^{2}\int_{\omega}e^{(N-1){\overline E}({\bf r})}\Pi_{i=2}^{N}e^{-E_{H}({\bf r},{\bf r}_{i})}d{\bf r}_{2}...d{\bf r}_{N}-\nonumber\\ -2(N-1)\sum_{i=2}^{N}\int_{\omega}\left[\int_{\Omega_{N-2}}e^{(N-2){\overline E}({\bf r})}\Pi_{l\neq i}e^{-E_{H}({\bf r},{\bf r}_{l})}d{\bf r}_{l}\right]e^{{\overline E}({\bf r})-E_{H}({\bf r},{\bf r}_{i})}\left[\nabla_{{\bf r}}{\overline E}({\bf r})\cdot \nabla_{{\bf r}}E_{H}({\bf r},{\bf r}_{i})\right]d{\bf r}_{i}
+\nonumber\\ +\sum_{i=2}^{N}\int_{\omega}\left[\int_{\Omega_{N-2}}e^{(N-2){\overline E}({\bf r})}\Pi_{l\neq i}e^{-E_{H}({\bf r},{\bf r}_{l})}d{\bf r}_{l}\right]e^{{\overline E}({\bf r})-E_{H}({\bf r},{\bf r}_{i})}\left|\nabla_{{\bf r}}E_{H}({\bf r},{\bf r}_{i})\right|^{2}d{\bf r}_{i}
+ \nonumber\\ +
\sum_{i}\sum_{j\neq i}\int_{\omega}\int_{\omega}\left[\int_{\Omega_{N-3}}e^{(N-3){\overline E}({\bf r})}\Pi_{l\neq i; l\neq j}e^{-E({\bf r},{\bf r}_{l})}d{\bf r}_{l}\right]e^{{\overline E}({\bf r})-E({\bf r},{\bf r}_{i})}e^{{\overline E}({\bf r})-E({\bf r},{\bf r}_{j})}\times\nonumber\\ \times\left|\nabla_{{\bf r}}E({\bf r},{\bf r}_{i})\cdot\nabla_{{\bf r}}E({\bf r},{\bf r}_{j})\right|d{\bf r}_{i}d{\bf r}_{j}.
\label{big1}
\end{eqnarray}
Using the  indistinguishability, taking into account that all the expressions of the form $\int e^{{\overline E}({\bf r})-E({\bf r},{\bf r}_{l})}d{\bf r}_{l}=1; \forall {\bf r}$ and factorizing all the ${\bf r}_{l}$'s that do not contain the variable ${\bf r}_{i}$ present in $\nabla_{{\bf r}}E({\bf r},{\bf r}_{i})$, we obtain:
\begin{eqnarray}
I({\bf r})=(N-1)^{2}\left|{\overline E}({\bf r})\right|^{2}-2(N-1)^{2}\int_{\omega}e^{{\overline E}({\bf r})-E({\bf r},{\bf r'})}\nabla_{{\bf r}}{\overline E}({\bf r})\cdot\nabla_{{\bf r}}E({\bf r},{\bf r'})d{\bf r'}+ \nonumber\\ +(N-1)\int_{\omega}e^{{\overline E}({\bf r})-E({\bf r},{\bf r'})}\left|\nabla_{{\bf r}}E({\bf r},{\bf r'})\right|^{2}d{\bf r'}+ \nonumber\\ +(N-1)(N-2)\int_{\omega}\int_{\omega}e^{{\overline E}({\bf r})-E({\bf r},{\bf r'})}e^{{\overline E}({\bf r})-E({\bf r},{\bf r''})}\nabla_{{\bf r}}E({\bf r},{\bf r'})\cdot\nabla_{{\bf r}}E({\bf r},{\bf r''}) d{\bf r'}d{\bf r''}.
\label{simpl}
\end{eqnarray}
The last term on the r.h.s. of Eq.\ref{simpl},characterized by a sort of double independent non locality, (${\bf r'}$ and ${\bf r''}$),
can be reduced to a simple non locality since:
\begin{equation}
\nabla_{{\bf r}}E({\bf r},{\bf r'})\cdot\nabla_{{\bf r}}E({\bf r},{\bf r''})=\partial_{x}E({\bf r},{\bf r'})\partial_{x}E({\bf r},{\bf r''})+\partial_{y}E({\bf r},{\bf r'})\partial_{y}E({\bf r},{\bf r''})+
\partial_{z}E({\bf r},{\bf r'})\partial_{z}E({\bf r},{\bf r''})
\label{partial}
\end{equation}
thus:
\begin{eqnarray}
(N-1)(N-2)\int_{\omega}\int_{\omega}e^{{\overline E}({\bf r})-E({\bf r},{\bf r'})}e^{{\overline E}({\bf r})-E({\bf r},{\bf r''})}\nabla_{{\bf r}}E({\bf r},{\bf r'})\cdot\nabla_{{\bf r}}E({\bf r},{\bf r''}) d{\bf r'}d{\bf r''}=\nonumber\\ =(N-1)(N-2)e^{2{\overline E}({\bf r})}\sum_{k=1,3}\int_{\omega}e^{E({\bf r},{\bf r'})}\partial_{x_{k}}E({\bf r},{\bf r'})d{\bf r'}\int_{\omega}e^{E({\bf r},{\bf r''})}\partial_{x_{k}}E({\bf r},{\bf r''})d{\bf r''}
\label{prmfin}
\end{eqnarray}
 where $x_{1}=x; x_{2}=y; x_{3}=z$, and from Eq.\ref{prmfin}, since $\partial_{x_{k}}E({\bf r},{\bf r'})$ and $\partial_{x_{k}}E({\bf r},{\bf r''})$ are independent from each other, one obtains:
\begin{equation}
e^{2{\overline E}({\bf r})}\sum_{k=1,3}\int_{\omega}e^{E({\bf r},{\bf r'})}\partial_{x_{k}}E({\bf r},{\bf r'})d{\bf r'}\int_{\omega}e^{E({\bf r},{\bf r''})}\partial_{x_{k}}E({\bf r},{\bf r''})d{\bf r''}=e^{2{\overline E}({\bf r})}\sum_{k=1,3}\left[\int_{\omega}e^{E({\bf r},{\bf r'})}\partial_{x_{k}}E({\bf r},{\bf r'})d{\bf r'}\right]^{2}.
\label{squad}
\end{equation}  
Inserting Eq.\ref{squad} in Eq.\ref{simpl}, we obtain an analytic expression for $I({\bf r})$, since 
all the terms involved in Eq.\ref{simpl} and Eq.\ref{squad} are known explicitly:
\begin{equation}
E_{H}({\bf r},{\bf r'})=\frac{\rho({\bf r})\rho({\bf r'})}{|{\bf r}-{\bf
    r'}|}
\end{equation}
and;
\begin{equation}
{\overline E}({\bf r})=-log\left[\int_{\omega}e^{-E_{H}({\bf r},{\bf
      r'})}d{\bf r'}\right]
\end{equation}
The physical interpretation of the results obtained is the following,
the non local Fisher information functional or kinetic correlation functional, can be well described by the classical Coulomb interaction
between the electrons. Such a conclusion is based on the fact that the
exponential form is physically ''reasonable'', however the related
mathematical prescriptions do exclude that the exchange and correlation
interaction plays a role in this case. Rather, it leads to the
conclusion that the classical Coulomb interaction naturally 
dominates the kinetic correlation effects since it perfectly matches the
physical arguments and the mathematical requirements.
\subsection{Discussion and Conclusions}
The proposed exponential form for the
conditional probability has been obtained on the basis of well founded physical
arguments and rigorous mathematical prescriptions. Interestingly, starting
from a rather different point of view and aiming at a different goal,
Ceperly\cite{cep} and Jastrow \cite{jast}, choose, on the basis of physical
intuition, the correlation part of a fermionic wavefunction as $e^{-u(|{\bf
    r}_{i}-{\bf r}_{j}|)}$, i.e. a repulsive pseudopotential. Moreover, Wigner
and Seitz \cite{wigseit}, and Wigner \cite{wig}, conclude that the conditional
probability part contained into a fermionic wavefunction, must be of the form
$f(|{\bf r}_{i}-{\bf r}_{j}|)$. 
This suggest
that indeed, the physical principle we have used, that is a high Coulomb interaction
corresponds to a less favorable situation compared to the situation of  a
weaker interaction, is a justified argument. Moreover, the exponential form given to the conditional probability,
which resembles the behavior just explained and fulfills all the mathematical
requirements, although not exact, cannot be far from the exact one
, and thus, at least qualitatively,
also the functional obtained is reasonably rigorous. A practical advantage of this approach is that it is not difficult, in principle, to implement this idea in a self-consistent calculation for atoms with a statitically relevant number of electrons (e.g. metals), where the Thomas-Fermi and Weiszacker terms describes well the the local part of the kinetic energy. In this case one can use the spherical symmetry and then compare the results with those obtained using, for example, the Hartree-Fock method. This test will certainly show possible advantages or disadvantages of this approach and thus inspire new ideas to improve it.

\acknowledgments

We thank M.Deserno, R.A.Mosna and I.P.Hamilton for helpful comments.

\end{document}